**Title:**

# Classical limit of Rabi nutations in spins of ferromagnets

**Authors:**

Amir Capua[1], Charles Rettner[1], See-Hun Yang[1], Stuart S. P. Parkin[1,2]

**Affiliations:**

[1] IBM Research Division, Almaden Research Center, 650 Harry Rd., San Jose, California 95120, USA
[2] Max Planck Institute for Microstructure Physics, Halle (Saale), D-06120, Germany

**Abstract:**

Rabi oscillations describe the interaction of a two-level system with a rotating electromagnetic field. As such, they serve as the principle method for manipulating quantum bits. By using a combination of femtosecond laser pulses and microwave excitations, we have observed the classical form of Rabi nutations in a ferromagnetic system whose equations of motion mirror the case of a precessing quantum two-level system. Key to our experiments is the selection of a subset of spins that is in resonance with the microwave excitation and whose coherence time is thereby extended. Taking advantage of Gilbert damping, the relaxation times are further increased such that mode-locking takes place. The observation of such Rabi nutations is the first step towards potential applications based on phase-coherent spin manipulation in ferromagnets.

**Main Text:**

A practical gateway to the quantum world is provided by macroscopic quantum systems that are large cooperative ensembles (*1*). Superfluids (*2, 3*), superconductors (*4, 5*) and ultracold dilute atomic vapors (*6-8*), are examples of such systems. Another example are magnon gases in ordered ensembles of magnetic moments that form a macroscopic state where the quantum nature is unveiled even at room temperature (*9*). In ferromagnets, the macroscopic quantum behavior asserts itself at low temperatures (mK) and/or small enough length scales (nanometer) (*10-12*). In that limit the angular momentum observables obey the classical equations of motion. Hence, a great deal of insight into the quantum word is gained from studies of classical analogs (*13*).

Although isolated electron or nuclear spin states are ideal candidates for quantum information processing (*14-18*), the abundant spin states in ferromagnetic systems are not currently considered suitable for such applications. Their spin states lack protection due to spin-spin and spin-lattice interactions. While these may be overcome in the quantum regime (*10*), their coherent manipulation remains unexplored, even in the classical limit.

The initialization, manipulation, and readout of spin ensembles in ferromagnetic systems requires operation in the non-adiabatic regime. This regime pertains whenever an oscillatory field and the state representing the ensemble are not in equilibrium. The adiabatic interaction has been primarily explored using ferromagnetic resonance (FMR) methods where steady state spin precessions are driven continuously by an oscillatory microwave field. Similar studies of the magnetic order have also been conducted by studying the impulse responses in the absence of the rotating field using the time-resolved magneto-optical Kerr effect (TR-MOKE) (*19-23*). In this technique the free induction decay response of a ferromagnet is triggered by an intense optical pulse which disturbs the magnetic order and drives the system away from the equilibrium state. Despite the extensive studies of spin dynamics in ferromagnetic metals, little attention has been given to non-adiabatic transitions. This mode of operation can be accessed whenever the driven spin precessions are disrupted and is achieved by either modifying the state of the oscillatory magnetic field, or that of the magnetization. While the former method is more commonly used, for instance by applying a $\pi$-pulse (*24*), the latter is adopted here.

Using ultrashort optical pulses to perturb a microwave driven ferromagnetic systems (*19*), we study the non-adiabatic regime and show that Rabi nutations in their classical form can be revealed in a ferromagnet. We observe a chirping of the precession frequency, and study the ability to manipulate the spin states in the presence of significant inhomogeneous broadening. In agreement with Gilbert's damping theory (*25*), the intrinsic relaxation times, which represent the loss of spin angular momentum to the environment, can be extended by tuning the external magnetic field (*26*). In such cases, consecutive optical pulses act to synchronize the phases of the precessing spins (*27*). Consequently, spin-mode-locking is initiated having the form of intense pulsations of the magnetization. Our experiments reveal that the microwave signal induces coherence in the ensemble by selecting a subset of spins that are driven resonantly with the field. Hence, the ensemble dephasing is suppressed and relaxation times that represent more closely those of individual spins result.

The sample studied was a $Co_{36}Fe_{44}B_{20}$ film with a thickness of 11 Å that was perpendicularly magnetized and grown by magnetron sputtering. The effective anisotropy field, $\mu_0 H_{Keff}$, was measured to be ~ 140 mT, with $\mu_0$ being the magnetic permeability. From TR-MOKE measurements of the free induction decay responses as a function of applied field (Fig. 1A) a Gilbert damping constant, $\alpha$, of 0.023 and a distribution of the effective anisotropy field, $\mu_0 \Delta H_{Keff}$, of 17.5 mT were determined (*28, 29*). $\alpha$ represents the losses of spin angular momentum without the ensemble dephasing while $\Delta H_{Keff}$ allows to determine the inhomogeneous broadening of the resonance linewidth (*28*).

The basic concept of our experiment is presented in Fig. 1B. A microwave field is used to drive spin precessions in the film which are then perturbed by a femtosecond optical pulse while being phase locked with the microwave oscillator (*30*). The temporal recovery is recorded by a weak optical pulse probe as a function of a pump-probe delay time. The external magnetic field was applied in the sample plane causing precessions to occur about the *x*-axis (see figure) while the out-of-plane component of the magnetization, $m_z$, was detected in a polar-MOKE configuration (*28*).

A measurement of the non-adiabatic interaction is presented in Fig. 1C for three values of externally applied field, $H_0$, and a microwave frequency of 10 GHz. The resonance field, $H_{res}$, corresponding to this frequency is $\mu_0 H_{res}$ ~ 450 mT (Fig. S1). It is seen that a distinct envelope modulates the 10 GHz oscillations of the precessional motion. Furthermore, this envelope exhibits a systematic behavior; the time of its minimum, as indicated by the arrows in the figure, increases as $H_0$ approaches $H_{res}$.

The responses for a complete set of applied fields, are illustrated in Fig. 1D. At $H_0 < H_{res}$, the shift in time of the minima is seen clearly and forms a "valley". At $H_0 > H_{res}$, a maximum is initially formed instead, making the response *asymmetric*. Given that the conditions for the non-adiabatic interaction prevail, the related times of these signatures, for example the time of the minima, $T$, should be describable by the inverse of the generalized Rabi frequency. In the absence of the magnetocrystalline and demagnetization fields, $T$ is given by:

$$T = \frac{2\pi}{\sqrt{\left(\gamma\mu_0 H_0 - \omega_{rf}\right)^2 + \left(\gamma\mu_0 h_{rf}\right)^2}} \qquad (1)$$

Here, $\gamma$, $h_{rf}$, and $\omega_{rf}$ are the gyromagnetic ratio, microwave amplitude, and microwave angular frequency, respectively. The times obtained by the Rabi formula are overlaid on the measured responses of Fig. 1D and are seen to agree well with the observed signatures, demonstrating Rabi nutations in a ferromagnet in the classical limit. Also the second nutation is readily seen. It is instructive to notice that the microwave amplitude is only $\mu_0 h_{rf} \sim 0.8$ mT in these measurements. Hence, the main contribution to the minima time, $T$, stems from the off-resonance term, $\gamma\mu_0 H_0 - \omega_{rf}$.

The field dependent phase responses also reveal intricate details of the dynamics and are analyzed by plotting the dataset of Fig. 1D as a two-dimensional contour plot (Fig. 2A). Before the pump pulse arrives, a net phase shift of ~ 0.75 π is measured across the resonance (enlarged in Fig. 2B). This phase shift is smaller than the expected theoretical value of π and is related to the relatively large Gilbert damping. Surprisingly, at times (pump-probe delays) well after the perturbation, a phase shift of ~ 2.75 π is observed as $H_0$ is varied (Fig. 2C). To understand the origin of this behavior we analyze the instantaneous frequency profiles (Fig. 2D). Apart from the sharp transient at $t = 0$, we extract negative, zero, and positive chirp profiles corresponding to $H_0 < H_{res}$, $H_0 = H_{res}$, and $H_0 > H_{res}$, respectively. At long delays, the instantaneous frequency recovers to the driving frequency, independent of $H_0$. This behavior is explained qualitatively by recalling that Rabi oscillations can be regarded as a beating of the natural transient response of the system at the angular frequency of $\gamma\mu_0 H_0$ with the steady state response at $\omega_{rf}$ (31). Hence, when $\gamma\mu_0 H_0 < \omega_{rf}$, a negative chirp initially takes place which recovers to $\omega_{rf}$. The same explanation holds also for other $H_0$ values. This reasoning also accounts for the asymmetry seen in the responses of Fig. 1D. As $H_0$ varies, the pump pulse perturbs the magnetization at different points along the precession trajectory owing to the phase shift associated with the resonance. The resultant beating response then changes from a destructive nature to a constructive interaction. Therefore, variation of $H_0$ provides a means of controlling the effective "area" (the time-integrated Rabi frequency) of the microwave radiation. A theoretical description of the interaction using the Landau-Lifshitz-Gilbert equation is further discussed in the supplementary materials section.

An important aspect of Rabi nutations is the dependence of their frequency on the magnitude of the microwave field. This dependence is most readily seen under resonance conditions in which case the angular Rabi frequency simplifies to $\Omega_r = \gamma \mu_0 h_{rf}$. The measured results are shown in Fig. 3A. In contrast to our expectations, no dependence of the envelope on the amplitude of the microwave is revealed. This apparent discrepancy is resolved by considering the contributions to Eq. (1). The maximal applied microwave field amplitude was $\mu_0 h_{rf} \sim 7.5$ mT while the inhomogeneous linewidth broadening at 10 GHz, as derived from the value of $\Delta H_{Keff}$, is 10.5 mT (*28*). Hence, the detuning term in Eq. (1) is still significant so that the Rabi frequency is mainly determined by the off-resonant contribution rather than by the microwave power. In order to observe the dependence on the microwave power, the contribution of the inhomogeneous broadening must first be suppressed. This was achieved by repeating the measurements on a single crystal sample, in the form of a 4 nm thick epitaxially grown Fe film. In contrast to the sputter deposited film, the envelope exhibits a clear dependence on the microwave amplitude (Fig. 3B). The expected increase in the associated time scales describing the envelope is seen for increasing amplitudes as predicted by Rabi's formula.

Next, we turn to show that the train of optical pulses can synchronize the phases of the spin precessions and induce pulsations of the magnetization, namely, spin mode-locking. This mode of operation can be reached if the responses generated by subsequent pump pulses of the pulse train interfere. Therefore, this regime requires that the transient part of the responses persist for a duration longer than the laser repetition time, $T_R$ (Fig. 4A).

As follows from Gilbert's theory for damping, the rate of transfer of spin angular momentum to the lattice can be controlled by the magnitude of $H_0$. This process is quantified using the intrinsic relaxation time, $\tau_{int}$, and is given by $(\gamma\mu_0 H_0 \alpha)^{-1}$ in the limit where only the externally applied field is present. Accordingly, interference effects are expected at low $H_0$ values. The nature of the interference will then depend on the arrival time of the optical pump pulse within the microwave cycle. This time is represented by $\Phi$ which is the relative phase between the microwave signal and the pump pulse (Fig. 4A).

The measured responses as a function of $\Phi$ are presented in Figs. 4B & 4C. At high magnetic field ($\mu_0 H_0$ = 450 mT) and short $\tau_{int}$ (~ 1.1 ns) (*28*) compared to the laser repetition time, $T_R$, of 12.5 ns, the interaction of each pump pulse within the train of pulses can be regarded as an isolated event (Fig. 4B). Variation of $\Phi$ has no effect on the envelope; the carrier wave simply shifts within the same envelope. In contrast, for low external magnetic fields ($\mu_0 H_0$ = 90 mT) and correspondingly long $\tau_{int}$ (~ 5 ns), interference occurs and the moment at which the optical pulse is sent becomes critical (Fig. 4C). For $\Phi = 90°$, constructive interference results in a sharp pulsation of the magnetization. Likewise, for additional 180°, at $\Phi = 270°$, pulsations of opposite polarity are generated. However, when the phase is tuned to $\Phi = 0°$ and $\Phi = 180°$, destructive interference takes place and no pulsations are observed. The existence of the pulsations indicates that the spins have become synchronized, i.e., mode-locking takes place (*27*).

In addition to the intrinsic relaxation, the decay of the transient response is governed also by the dephasing of the inhomogeneously broadened ensemble. This process is represented by the ensemble dephasing time, $\tau_{IH}$, so that the effective decay time of the magnetization of the entire ensemble, $\tau_{eff}$, is given by: $1/\tau_{eff} = 1/\tau_{int} + 1/\tau_{IH}$ (*28, 29, 32*).

Interestingly, while a fundamentally different dependence on $\Phi$ is observed in the two regimes of Figs. 4B and 4C, the inhomogeneous broadening causes $\tau_{eff}$ to be very similar in both cases and corresponds to ~ 0.51 ns and ~ 0.49 ns for $\mu_0 H_0 = 450$ mT and $\mu_0 H_0 = 90$ mT, respectively. This fact shows that the long intrinsic relaxation time, $\tau_{int}$, in the case of low $H_0$ (Fig. 4C) can be sensed despite the significant ensemble dephasing. By use of the relations $\Delta\omega_{int} = 2/\tau_{int}$ and $\Delta\omega_{IH} = 2/\tau_{IH}$ for the intrinsic resonance linewidth and inhomogeneous broadening, respectively, $\Delta\omega_{int} \approx 1.75$ rad·$GHz$ and $\Delta\omega_{IH} \approx 2.15$ rad·$GHz$ were extracted for $\mu_0 H_0 = 450$ mT, while $\Delta\omega_{int} \approx 0.43$ rad·$GHz$ and $\Delta\omega_{IH} \approx 3.66$ rad·$GHz$ were found for $\mu_0 H_0 = 90$ mT. These linewidths are illustrated in the lower schematic of Figs. 4B & 4C. In contrast to the high $H_0$ case, at low $H_0$ only a subset of spins which exhibit long $\tau_{int}$ are interacting, namely, the microwave induces coherence in the ensemble. The action of the oscillatory field is to stimulate the subset of spins that are driven resonantly, while suppressing the off-resonance subsets. Hence, the inhomogeneity is overcome and $\tau_{eff}$ extends towards its upper limit of $\tau_{int}$.

The action of "filtering" by the microwave signal is further emphasized by examining the free induction decay responses. Similar long intrinsic relaxation times that are responsible for the mode-locking with $\mu_0 H_0 = 90$ mT, also dominate the corresponding TR-MOKE measurement at $\mu_0 H_0 = 100$ mT (Fig. 1A), for example. In contrast to the microwave driven measurement, this response shows that the ensemble dephases within $T_R$ and exhibits no signs of coherent interference, as apparent from times around $t = 0$ (*27*). Furthermore, a closer inspection of the measurements at the high $H_0$ of Fig. 1D also turn out to reveal slight signatures of extended coherence that last for the duration of $T_R$, and are clearly not found in the corresponding TR-MOKE responses. These signatures are discussed in the supplementary materials section and once more demonstrate the ability of our technique to observe details that are obscured by the ensemble dephasing.

Lastly, in the high magnetic field limit of Fig. 4B, variation of $\Phi$ was shown to cause no effect on the envelope. This observation seemingly contradicts the explanation behind the appearance of the asymmetric signature of Fig. 1D which was attributed to the phase shift associated with the resonance. However, a fundamental difference exists between the two measurements; in the former case, the phase-shift stems from delaying the microwave signal with respect to the optical pulses while in the latter it stems from the phase response of the resonance.

In summary, we have demonstrated a technique that has revealed the classical form of Rabi nutations in a ferromagnetic system. Our experiments show that a "purified" sub-ensemble is generated and whose dephasing is largely reduced. Extensions of the present work include studies of the quantum regime in mesoscopic ferromagnetic structures, as well as studies of more complex dynamics such as the spin-Hall effect or the spin transfer torque in the non-adiabatic regime. Connecting these effects and other spintronic effects with quantum processing may prove to be useful in overcoming problems that hinder even some mature quantum information technologies (*33*).

**Acknowledgments:**

We thank Dr. Dan Rugar, Dr. Chris Lutz and Dr. John Mamin for fruitful discussions and Chris Lada for expert technical assistance. A.C. thanks the Viterbi Foundation and the Feder Family Foundation for supporting this research.

**References and Notes:**

1. A. J. Leggett *et al.*, Dynamics of the dissipative two-state system. *Reviews of Modern Physics* **59**, 1 (1987).
2. D. J. Bishop, J. D. Reppy, Study of the Superfluid Transition in Two-Dimensional 4^He Films. *Physical Review Letters* **40**, 1727 (1978).
3. R. Desbuquois *et al.*, Superfluid behaviour of a two-dimensional Bose gas. *Nat Phys* **8**, 645 (2012).
4. J. Clarke, A. N. Cleland, M. H. Devoret, D. Esteve, J. M. Martinis, Quantum Mechanics of a Macroscopic Variable: The Phase Difference of a Josephson Junction. *Science* **239**, 992 (1988).
5. A. J. Berkley *et al.*, Entangled Macroscopic Quantum States in Two Superconducting Qubits. *Science* **300**, 1548 (2003).
6. M. H. Anderson, J. R. Ensher, M. R. Matthews, C. E. Wieman, E. A. Cornell, Observation of Bose-Einstein Condensation in a Dilute Atomic Vapor. *Science* **269**, 198 (1995).
7. I. Bloch, T. W. Hansch, T. Esslinger, Measurement of the spatial coherence of a trapped Bose gas at the phase transition. *Nature* **403**, 166 (2000).
8. C. I. Hancox, S. C. Doret, M. T. Hummon, L. Luo, J. M. Doyle, Magnetic trapping of rare-earth atoms at millikelvin temperatures. *Nature* **431**, 281 (2004).
9. S. O. Demokritov *et al.*, Bose-Einstein condensation of quasi-equilibrium magnons at room temperature under pumping. *Nature* **443**, 430 (2006).
10. D. D. Awschalom, D. P. Divincenzo, J. F. Smyth, Macroscopic quantum effects in nanometer-scale magnets. *Science* **258**, 414 (Oct 16, 1992).
11. S. Kleff, J. v. Delft, M. M. Deshmukh, D. C. Ralph, Model for ferromagnetic nanograins with discrete electronic states. *Physical Review B* **64**, 220401 (2001).
12. M. M. Deshmukh *et al.*, Magnetic Anisotropy Variations and Nonequilibrium Tunneling in a Cobalt Nanoparticle. *Physical Review Letters* **87**, 226801 (2001).
13. A. Abragam, *The Principles of Nuclear Magnetism*. (Clarendon Press, Oxford 1961).
14. G. D. Fuchs, V. V. Dobrovitski, D. M. Toyli, F. J. Heremans, D. D. Awschalom, Gigahertz Dynamics of a Strongly Driven Single Quantum Spin. *Science* **326**, 1520 (2009).
15. P. Neumann *et al.*, Quantum register based on coupled electron spins in a room-temperature solid. *Nat Phys* **6**, 249 (2010).
16. E. A. Chekhovich *et al.*, Nuclear spin effects in semiconductor quantum dots. *Nat Mater* **12**, 494 (2013).
17. D. J. Christle *et al.*, Isolated electron spins in silicon carbide with millisecond coherence times. *Nat Mater* **14**, 160 (2015).
18. C. G. Yale *et al.*, Optical manipulation of the Berry phase in a solid-state spin qubit. *Nat Photon* **10**, 184 (2016).

19. E. Beaurepaire, J. C. Merle, A. Daunois, J. Y. Bigot, Ultrafast Spin Dynamics in Ferromagnetic Nickel. *Physical Review Letters* **76**, 4250 (1996).
20. B. Koopmans, M. van Kampen, J. T. Kohlhepp, W. J. M. de Jonge, Ultrafast Magneto-Optics in Nickel: Magnetism or Optics? *Physical Review Letters* **85**, 844 (2000).
21. J.-Y. Bigot, M. Vomir, E. Beaurepaire, Coherent ultrafast magnetism induced by femtosecond laser pulses. *Nat Phys* **5**, 515 (2009).
22. B. Koopmans *et al.*, Explaining the paradoxical diversity of ultrafast laser-induced demagnetization. *Nat Mater* **9**, 259 (2010).
23. D. Rudolf *et al.*, Ultrafast magnetization enhancement in metallic multilayers driven by superdiffusive spin current. *Nat Commun* **3**, 1037 (2012).
24. D. J. Griffiths, *Introduction to Quantum Mechanics*. (Pearson Prentice Hall, Upper Saddle River, NJ, ed. 2nd ed., 2005).
25. T. L. Gilbert, A phenomenological theory of damping in ferromagnetic materials. *Magnetics, IEEE Transactions on* **40**, 3443 (2004).
26. R. Hanson, V. V. Dobrovitski, A. E. Feiguin, O. Gywat, D. D. Awschalom, Coherent Dynamics of a Single Spin Interacting with an Adjustable Spin Bath. *Science* **320**, 352 (2008-04-18 00:00:00, 2008).
27. A. Greilich *et al.*, Mode locking of electron spin coherences in singly charged quantum dots. *Science* **313**, 341 (Jul 21, 2006).
28. Materials and methods are available as supplementary materials on Science Online.
29. A. Capua, S.-H. Yang, T. Phung, S. S. P. Parkin, Determination of intrinsic damping of perpendicularly magnetized ultrathin films from time-resolved precessional magnetization measurements. *Physical Review B* **92**, 224402 (2015).
30. I. Neudecker *et al.*, Modal spectrum of permalloy disks excited by in-plane magnetic fields. *Physical Review B* **73**, 134426 (2006).
31. H. C. Torrey, Transient Nutations in Nuclear Magnetic Resonance. *Physical Review* **76**, 1059 (1949).
32. S. Iihama *et al.*, Gilbert damping constants of Ta/CoFeB/MgO(Ta) thin films measured by optical detection of precessional magnetization dynamics. *Physical Review B* **89**, 174416 (2014).
33. L. M. K. Vandersypen *et al.*, Experimental realization of Shor's quantum factoring algorithm using nuclear magnetic resonance. *Nature* **414**, 883 (2001).

Fig. 1

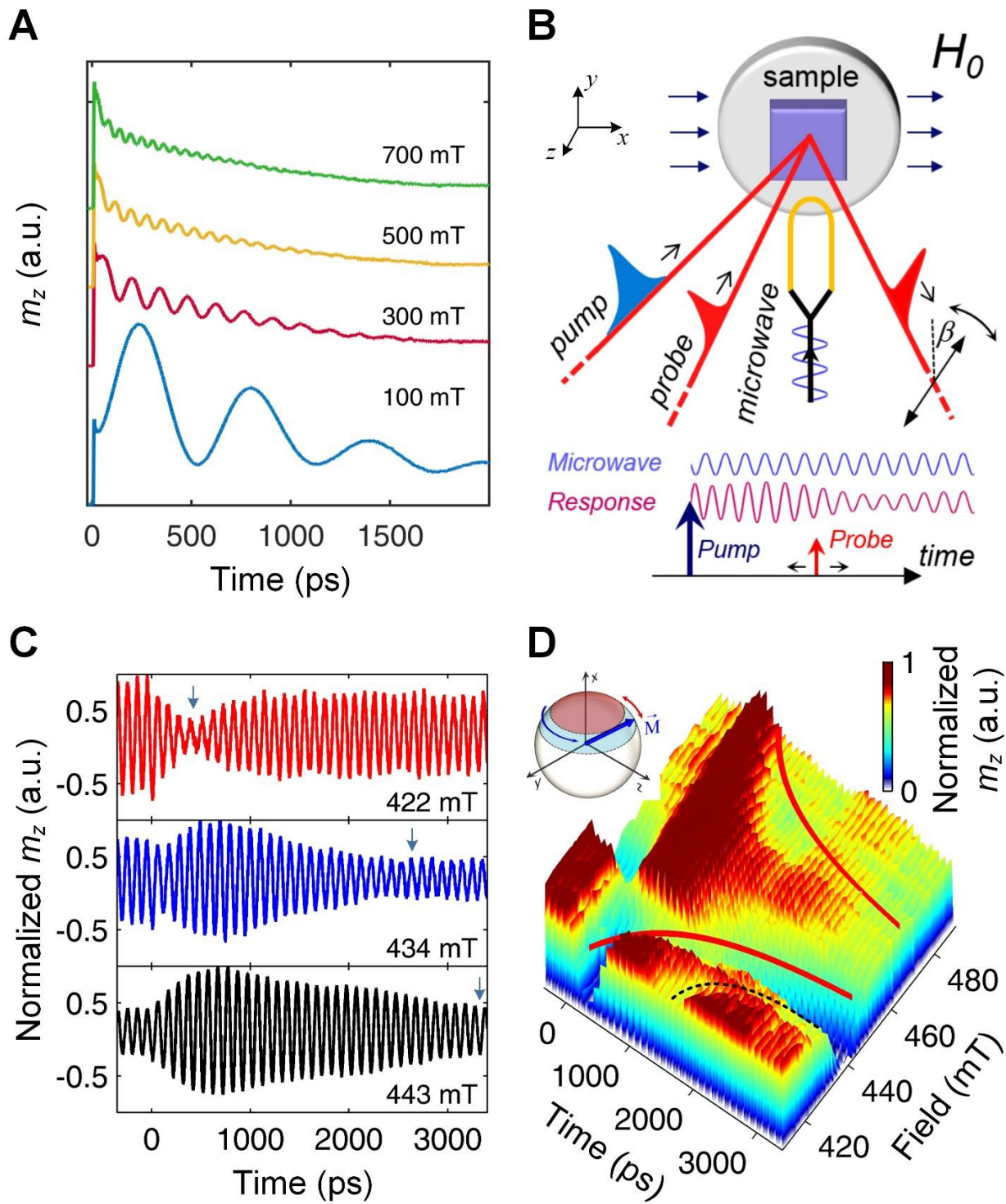

**Fig. 1. Measurement in the non-adiabatic regime.** (A) Free induction decay responses of a TR-MOKE experiment. Traces are shifted along the *y*-axis for clarity. (B) Schematic of the experimental setup. The magnetic film was patterned into a square island while the microwave signal was transmitted by a shorted Au microwire. The measured signal is proportional to the angle of polarization rotation, $\beta$, of the optical beam. Lower schematic shows the signals in time. (C) Temporal responses of the pump-probe ferromagnetic resonance measurement for three values of $H_0$ at 10 GHz and microwave field amplitude of ~ 0.8 mT. Each trace is normalized to the peak value. Arrows indicate the position of the minimum. The pump pulse arrives at $t = 0$ ps. (D) Measured temporal responses at 10 GHz for a complete range of applied fields. Each trace is normalized individually to the peak value. The solid red lines were plotted using the Rabi formula. Second oscillation is indicated by the guiding black dashed line. Inset illustrates the motion of the magnetization vector.

Fig. 2

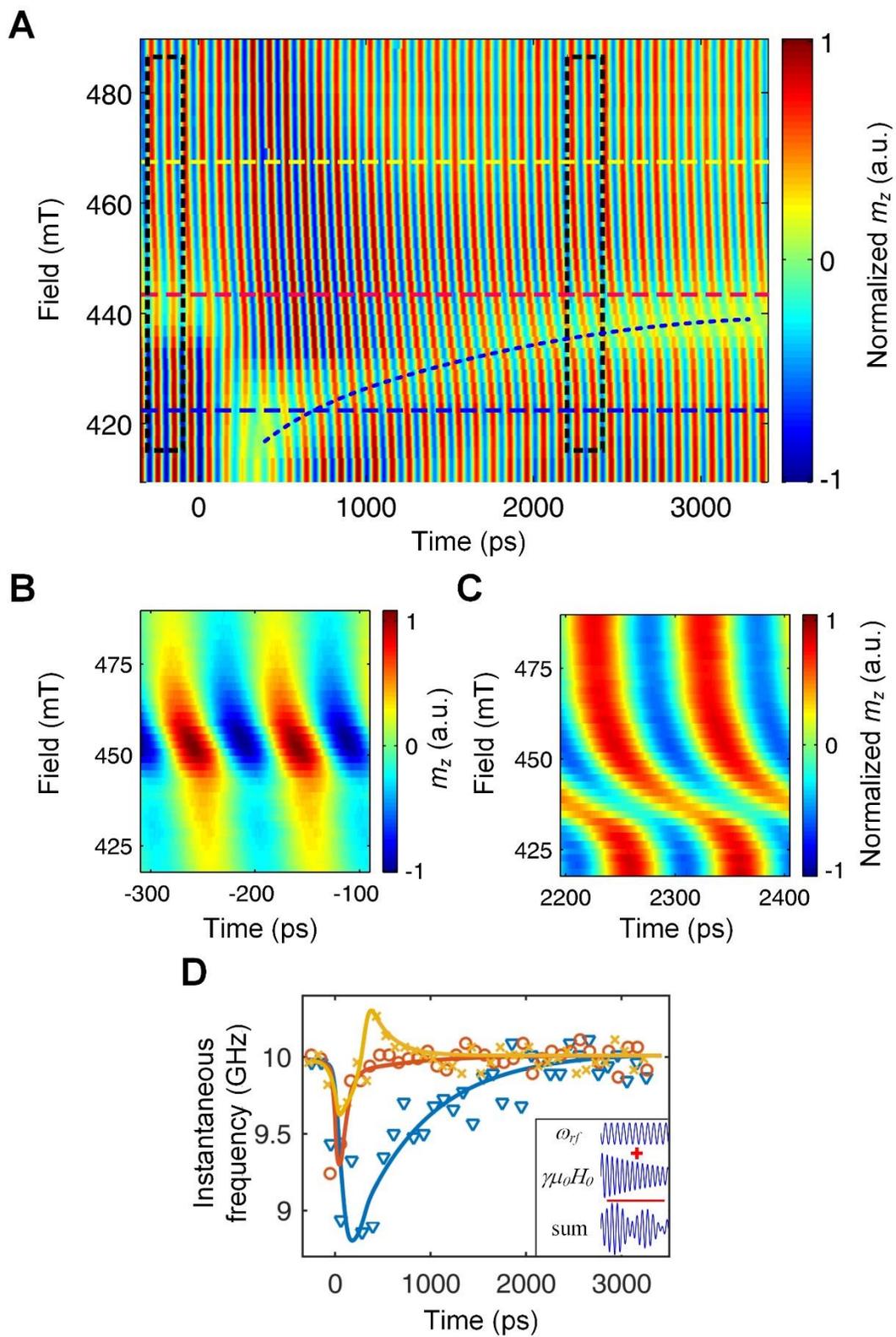

**Fig. 2. Phase response.** (A) Dependence of phase responses on the applied field at 10 GHz. Dataset of Fig. 1D is presented in a two-dimensional contour plot to represent the phase information. Each temporal response was normalized individually. Blue curved guiding line indicates the location of the "valley" in Fig. 1D. (B) Phase response prior to the perturbation. The figure presents a close-up of the black dashed area of panel (A) for times between -300 ps and -100 ps. An overall phase shift of ~ 0.75 $\pi$ is measured across the resonance. Data is not normalized. (C) Phase response at long delays corresponding to black dashed area in panel (A) which starts at 2200 ps. Data is presented in normalized units. The measured net phase shift across the resonance is ~ 2.75 $\pi$. (D) Instantaneous frequency profiles at $\mu_0 H_0$ values of 424 mT (blue), 444 mT (red), and 468 mT (yellow) corresponding the blue, red, and yellow dashed lines of panel (A), respectively. Inset illustrates a schematic of the beating of the steady state response of the system at $\omega_{rf}$ with the natural response at the angular frequency of $\gamma \mu_0 H_0$.

Fig. 3

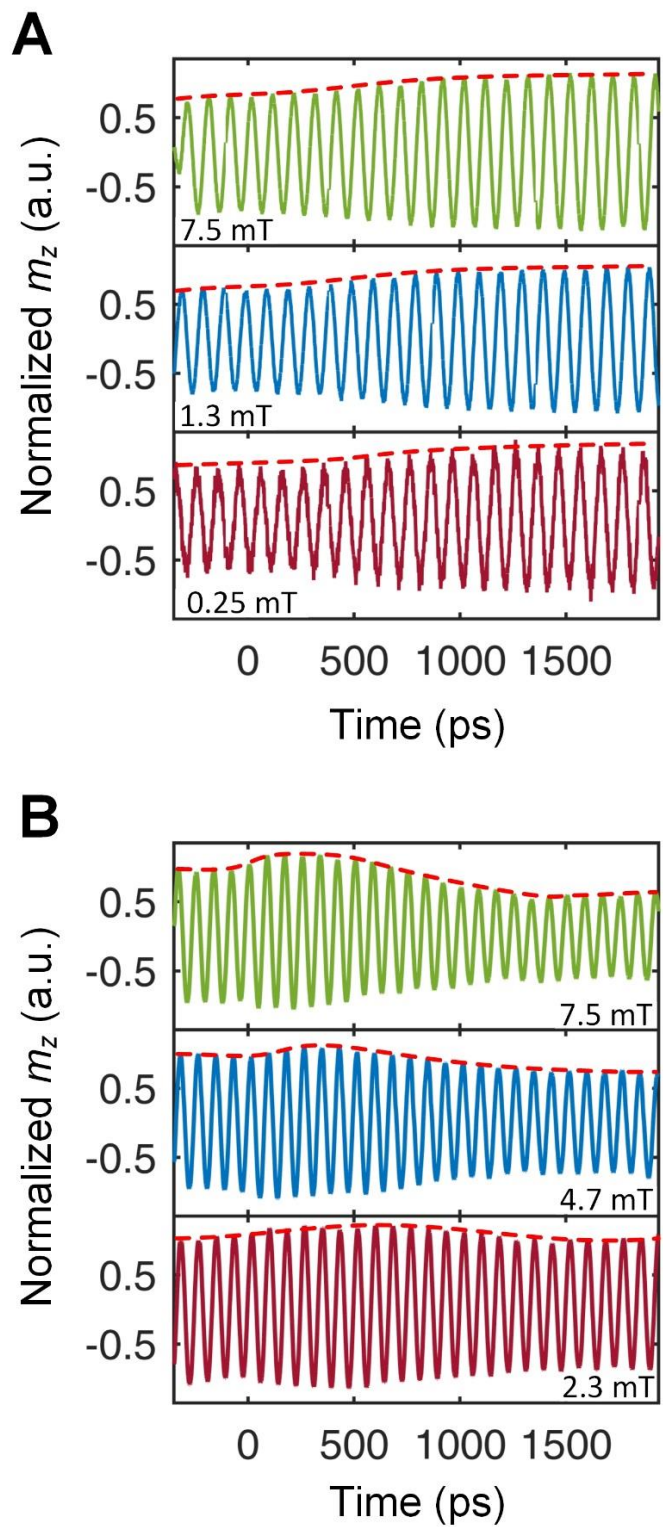

**Fig. 3. Microwave power dependence of the nutations.** (A) Temporal responses at various microwave field amplitudes for the CoFeB sample. Responses are presented for a frequency of 10 GHz and $\mu_0 H_0$ = 446 mT. (B) Temporal responses of the MBE grown Fe sample at 12 GHz and $\mu_0 H_0$ = 143 mT. The measurements in (A) and (B) were carried out at the resonance conditions. Envelopes of the responses are indicated by the guiding dashed lines.

Fig. 4

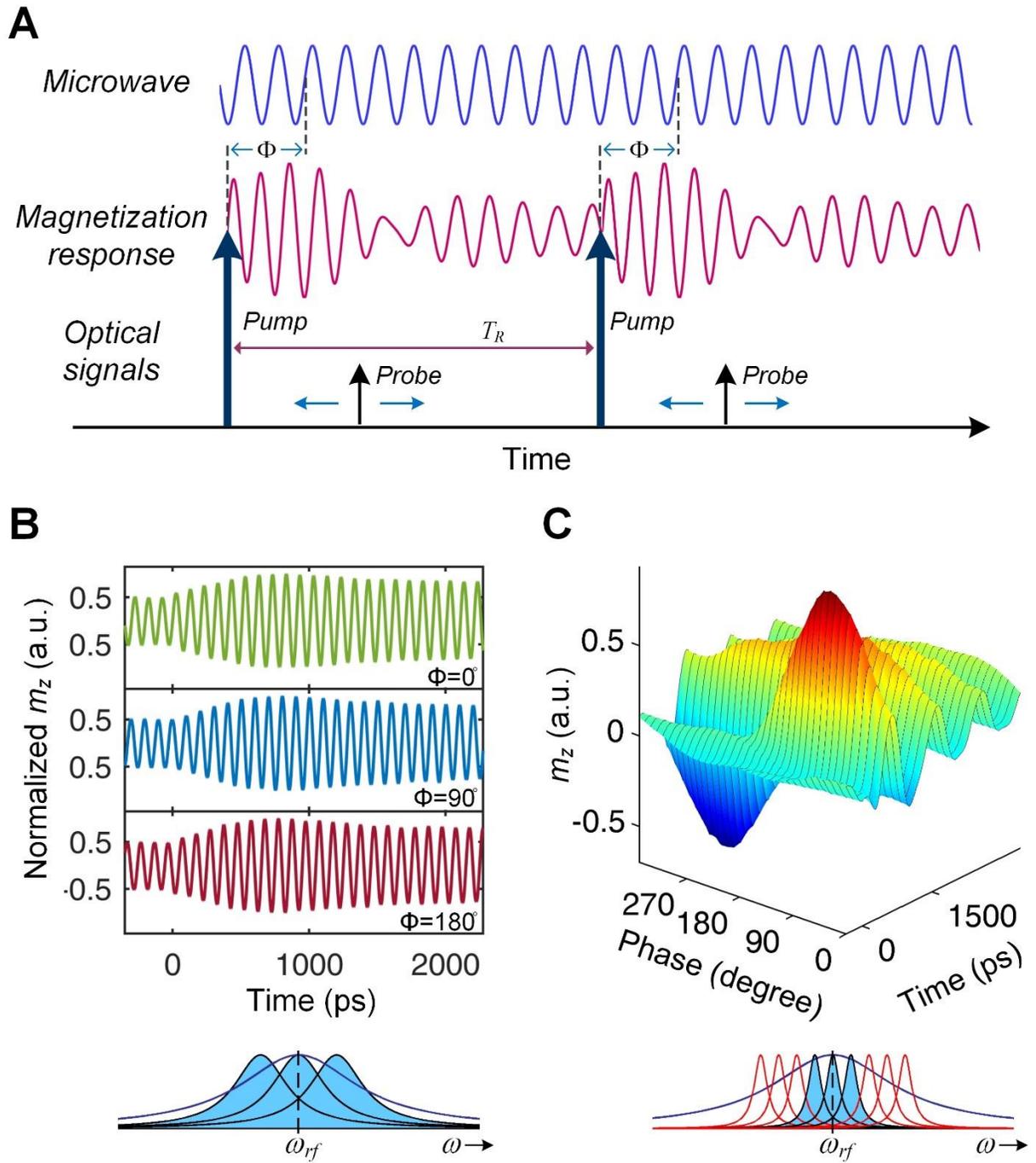

**Fig. 4. Phase dependent temporal responses for long and short intrinsic relaxation times.** (A) Schematic arrangement of signals in time. $T_R$ represents the laser repetition time. (B) Dependence of the temporal response on $\Phi$ for short $\tau_{int}$. Data presented for a frequency of 10 GHz and $\mu_0 H_0 = 450$ mT. The relative phase, $\Phi$, does not have a significant effect on the envelope. Similar behavior is recorded also at other bias fields (not shown). (C) Dependence of temporal response on $\Phi$ for long $\tau_{int}$. Data presented for a frequency of 1 GHz and $\mu_0 H_0 = 90$ mT. Data is shown for the CoFeB sample. Lower schematic is panels (B) and (C) illustrate the inhomogeneous broadening. Blue solid line indicates the total effective resonance linewidth which includes contributions of the inhomogeneous broadening. Shaded resonance linewidths indicate the subgroups that are selected by the microwave. Red solid lines indicate the subgroups that are not interacting with the microwave signal.

Supplementary Materials for

**Classical limit of Rabi nutations in spins of ferromagnets**

Amir Capua, Charles Rettner, See-Hun Yang, Stuart S. P. Parkin

**Materials and Methods**

Materials and Fabrication

The CoFeB film in this study was prepared on thermally oxidized Si(100) substrate and consisted of the following structure, starting from the substrate side: $SiO_2$ (250)/Ta (100)/CoFeB (11)/MgO (11)/Ta (30) (numbers are in nominal thicknesses in angstroms). The MgO layer was deposited by RF sputtering. The sample was annealed at a temperature of $275°C$ for 30 minutes while applying a 1 T field in the out-of-plane direction. In-plane and out-of-plane magnetization loops are shown in Fig. S1. The single crystalline 40 Å thick Fe film was grown on MgO(100) using molecular beam epitaxy.

For the pump-probe ferromagnetic resonance measurement the samples were patterned to a magnetic island of $20 \times 20$ $\mu m^2$ using electron-beam lithography. A shorted Au microwire serving as an RF-transmission line was patterned at a distance of 1 $\mu m$ away from the island by lift-off.

Ferromagnetic resonance pump-probe measurement

A Ti:Sapphire oscillator emitting ~ 70 fs pulses at 800 nm having energy of ~ 5 nJ per pulse was used for the optical measurements. The beam was focused to a spot size of approximately 10 $\mu m$. The probe pulses were attenuated by 20 dB relative to the pump. The time jitter between the optical pump and the microwave signal was measured to be smaller than 1 ps. All measurements were carried out at room temperature. The maximum microwave power applied was 1 W and corresponded to an amplitude of ~ 7.5 mT.

In the pump-probe ferromagnetic resonance measurements a double lock-in detection scheme was used by modulating the microwave signal at 50 KHz and the optical probe at 1 KHz. In order to exert sufficient torque by the optical pump, the external magnetic field was applied at an angle of $4°$ away from the sample plane. The same arrangement was applied also in the TR-MOKE measurements.

Numerical simulation

Calculation of the non-adiabatic interaction was carried out by numerically integrating the Landau-Lifshitz-Gilbert equation. Since the calculation does not account for the inhomogeneous broadening, it describes the experiment in a qualitative manner. In the calculation, the steady precessional state was first obtained before applying the perturbation. Two sources for the perturbation were introduced that gave the best results: quenching of the magnetization and introduction of a momentary phenomenological magnetic field. The latter was required in order to reproduce the phase response at positive times near $t = 0$, namely the curvature in the vertical contours of Fig. S2 appearing at times that immediately follow the pump. In Fig. S2, the wave fronts shift to later times as the field increases to a value of ~ 440 mT. When the field is further increased, the wave fronts shift to earlier times. Since the magnetization acquires a phase shift that is associated with the resonance at a field of about 450 mT, presenting an additional phenomenological magnetic field causes the magnetization to alter its motion. This additional torque was applied in the form of a 3 ps pulsed magnetic field of 60 mT which lied in the film plane orthogonal to the axis of precession. The recovery profile of the magnetization after quenching consisted of two time constants of 50 ps and 500 ps while the modulation depth was 5%.

The simulation result is shown in Fig. S3. Imprints of nutations on the amplitude of the precessions are readily seen. The formation of the "valley" as in Fig. 1D is also observed. This valley however appears also at field values which are larger than the resonance field in contrast to the measurement. At magnetic fields near resonance and immediately after zero time, similar contours to the ones shown in Fig. S2 are seen.

Extraction of decay times from TR-MOKE measurements

Extraction of the ensemble dephasing times and the intrinsic spin relaxation times from TR-MOKE measurements was based on the analysis presented in Ref. (*29*). Accordingly, $\alpha$ and $\Delta H_{Keff}$ were obtained by fitting the measured effective linewidths, $\Delta \omega_{eff}$, with the equation:

$$\Delta \omega_{eff} = \alpha \gamma \mu_0 \left(2H_0 - H_{Keff}\right) + \frac{\gamma H_0}{2\sqrt{H_0^2 - H_0 H_{Keff}}} \mu_0 \Delta H_{Keff} \quad \text{for} \quad H_0 > H_{keff}$$

$$\Delta \omega_{eff} = \alpha \gamma \mu_0 H_0 \left(\frac{2H_{Keff}}{H_0} - \frac{H_0}{H_{Keff}}\right) + \frac{\gamma H_{Keff}}{\sqrt{H_{Keff}^2 - H_0^2}} \mu_0 \Delta H_{Keff} \quad \text{for} \quad H_0 < H_{Keff}$$

Here $\Delta \omega_{eff} = 2/\tau_{eff}$, and $\tau_{eff}$ is the overall decay time of the precessional motion as measured in the TR-MOKE experiment. This analysis is valid for $H_0$ much larger or much smaller than $H_{Keff}$. The first terms of the equation represent the intrinsic linewidth $\Delta \omega_{int}$ while the second terms represent the inhomogeneous broadening $\Delta \omega_{IH}$. For $H_0 \sim H_{Keff}$, as for the case where $\mu_0 H_0 = 90$ mT, $\tau_{int}$ and $\tau_{IH}$ were extracted using a numerical method (*32*).

Signatures of extended coherence at high applied field and 10 GHz

Signatures of the extended coherences were even found in the phase responses at 10 GHz (Fig. 2A) where $\tau_{eff}$ and $\tau_{int}$ are shorter. At negative times, before the pump pulse arrives, the phase in Fig. 2B shifts slightly to negative values as $H_0$ increases to $\mu_0 H_0 \sim 450$ mT. This response differs from the responses measured when the optical pump is completely turned off (Fig. S4), indicating the slight remnant coherence from the previous cycle. Moreover, the negative phase shift resembles the behavior of the phases seen immediately after $t = 0$ (Fig. S2) implying a link between the responses manifested by long lasting coherence despite the short $\tau_{eff}$ and $\tau_{int}$ (*27*). A non-coherent process, such as a thermal process would have had an equal effect for all $H_0$ and would not have affected the phase in the manner observed. Once more, the corresponding TR-MOKE traces show no sign of the coherent interaction after $T_R$, demonstrating the ability to observe details that were obscured by the ensemble dephasing.

Figures

A

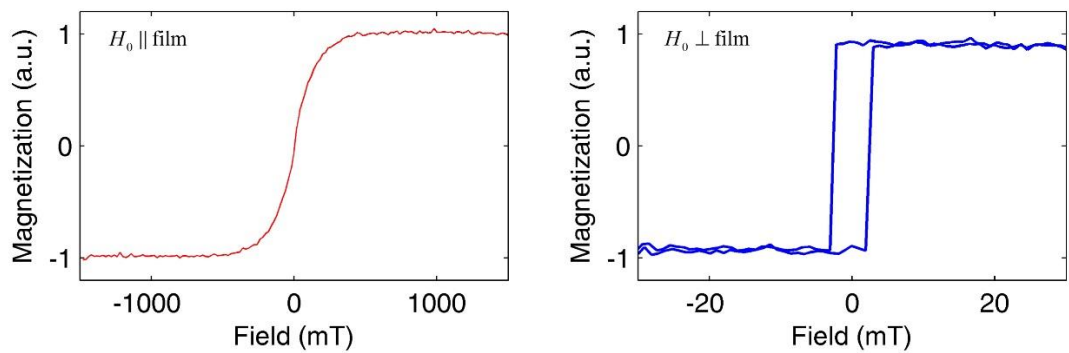

B

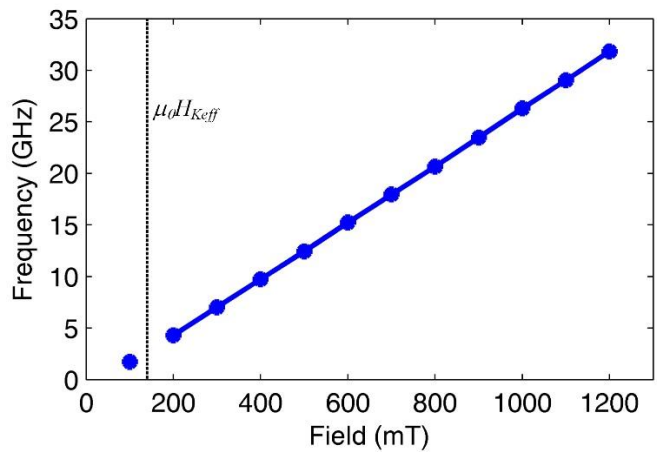

**Fig. S1.**
CoFeB film characterization. (A) In-plane and out-of-plane magnetization loops. (B) Frequency vs. applied field as measured in a TR-MOKE experiment. The magnetic field was applied at an angle of 4° away from the sample plane. Black dashed line indicates $H_{Keff}$.

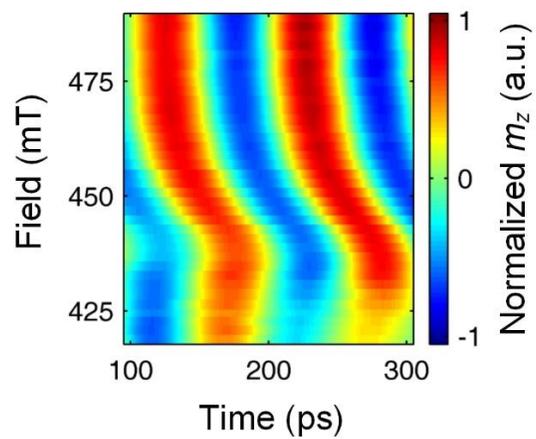

**Fig. S2**

Close-up of Fig. 2A for times between for times between 100 ps and 300 ps. A negative phase shift is seen as the field increases to a value of 440 mT after which a positive phase shift occurs when the field is further increased.

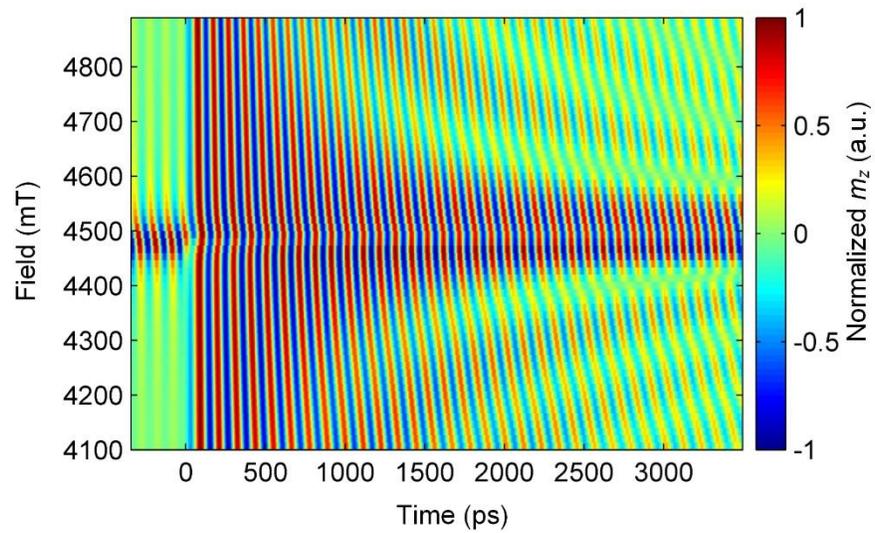

**Fig. S3**

Calculation of the out-of-plane component of the magnetization, $m_z$. The response at each bias field was normalized independently to reach a maximum value of unity.

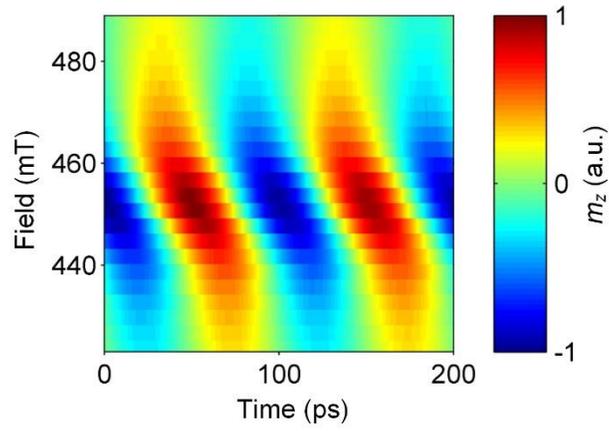

**Fig. S4**

Resonance response without the optical pump for the CoFeB sample. Measurement shows the out-of-plane component of the magnetization, $m_z$, at 10 GHz. The phase increases monotonically with the field in contrast to Fig. 2B of the main text. Data is not normalized.